\documentstyle[12pt,epsf,wrapfig]{article}
\begin{document}
\begin{flushright}
IHEP 98-21\\
\end{flushright}
\begin{center}
{\large\bf Model for three generations of fermions}\\
\vspace*{3mm}
V.V.~Kiselev\\
Institute for High Energy Physics,\\
Protvino, Moscow Region, 142284, Russia\\
E-mail: kiselev@mx.ihep.su, kiselev@th1.ihep.su, \\
Fax: +7-(095)-2302337, Phone: +7-(0967)-713780
\end{center}

\begin{abstract}
{A model is constructed for a chiral abelian gauge-interaction
of fermions and a potential of three higgses, so that the potential
possesses a discrete symmetry of the vacuum state, which provides
the introduction of three generations for the fermions. For the model
including charged currents, a matrix of the flavor mixing is
described, and a phase giving the violation of {\sf CP}-invariance
is determined. Some approximate relations connecting the values of mixing
matrix to the mass ratios for the fermions in different generations are
derived.}
\end{abstract}

\section*{Introduction}
One problem among those standing beyond the carefully studied Standard
Model of elementary particles \cite{ws} is the determination of an origin for
the fermion generations, which possess identical quantum numbers with respect
to gauge interactions and differ just in values of masses because of various
couplings to the Higgs fields \cite{hig}.

At present there are two approaches to this problem. According to the former,
the fermionic flavors are postulated as charges of a new ``horizontal'' gauge
interaction, an observation probability of which is suppressed at current
energies due to both small values of these charges and large masses of
corresponding intermediate bosons \cite{hor}.

In the letter, the appearance of flavors is caused by some dynamical
reasons. The following three ways are usually pointed out \cite{pecc}:

1. Dynamical origin of flavors. In the framework of models along this way,
fermions and Higgs fields are considered as composite states of fundamental
``preons'', so that the generations differ because of an internal preonic
structure \cite{composit}.

2. Random lattice model of space and gauge interactions \cite{random}. 
In such models the flavors appear as effective low-energy fields caused
by some nontrivial random interactions on the lattice in a deep region of the
Plank scale.

3. Geometric origin of generations: the superstring models with a
compactification of an extended-dimensions space \cite{GSW}. In this way the
flavors naturally appear due to an appropriate choice of compactified manifold
\cite{super}.

A common point of the mentioned ways is an essential extension for a set
of fundamental fields with respect to the Standard Model. A replication of
Higgs fields\footnote{In supersymmetric theories an additional higgs appears
even in the case of single generation because of the necessary introduction of
both a scalar field determining the masses of ``up'' kind electroweak fermions
at spontaneous breaking of the symmetry and a partner of the conjugated scalar
field determining the masses of ``down'' kind electroweak fermions
\cite{peskin}.} alike the number of generations seems to be a usual
peculiarity of such extensions. 

In the present paper we show how a model of potential for three Higgs fields
can be constructed in a manner which results in an appearance of vacuum state
possessing a discrete symmetry for higgses at spontaneous breaking of chiral
gauge symmetry, so that the discrete symmetry provides the introduction of
three generations of fermions. The advantage of such picture of generations
is the absence of direct postulation of ``preons'', random lattice dynamics or
superstring compactifications\footnote{Each of these ways has its own
disadvantage. So, the ``preonic'' way looks like a straightforward continuation
of the atomic idea for an embedding of more and more elementary particles
inside bigger nonfundamental objects, so that it is not clear whether a last
stage is reached or not, which is weakly attractive. In the random lattice
dynamics, it is problematic to build realistic models giving the fermion masses
close to the observed ones. In the superstring models a set of
compactifications is very broad, so that one can describe practically an
arbitrary situation in the low-energy region, but, unfortunately, one still
has not found a principal physical basis determining the parameters of the
compacification.}. Therefore, the offered model can be considered as quite a
self-consistent description for the introduction of fermion generations.
On the other hand, in the framework of mentioned ways for the construction of
fermion flavors, the model can be taken as a common low-energy limit for
three effective higgses interacting with the fermion field.

The higgs potential under consideration possesses a discrete symmetry of
three vacuum configurations which are determined by a minimum of energy for
constant fields. In accordance with transformations of this symmetry, a state
of the minimum can be obtained from another by a discrete change of complex
phases for the vacuum expectation values ({\sc vev}) of higgses. These
transformations are the representation of discrete group $Z_3$, whereas the
Higgs fields have the charges $\{-1, 0, 1\}$. The physical vacuum invariant by
the action of discrete group can be constructed as a direct product of minimal
energy states for the higgses. Then, the fermionic fields over each of the
three minima have different mass values along with the identical properties
with respect to the gauge interaction, which means the introduction of three
generations of fermions. In a matrix representation of the direct product for
the vacuum configurations of higgses, the mass matrix of fermions has a
symmetric texture determined by the discrete group. For the presence of charged
currents the representation of mass matrix can be fixed by the set of physical
parameters in the model: The products of higgs {\sc vev} by three Yukawa
coupling constants for fermions, whereas the only coupling constant is a
complex one. The necessary condition for the fixing of the mass matrix is the
introduction of complex phase causing the violation of {\sf CP}-invariance. In
real situation, when the fermion masses of major generation are much greater
than the masses of the junior ones, the relation of elements for the matrix of
charged current mixing \cite{ckm} with the physical parameters of model is
significantly simplified, so that up to small corrections over the mass ratios
of generations one can determine the phase of {\sf CP}-violation, and the
values of mixing angles are expressed through the mass ratios in the
parameterization recently introduced by Fritzsch and Xing \cite{fx} and also
discussed in \cite{rasin} \footnote{The author believes that the classification
of parameterizations for the mixing matrix was performed independently by
Fritzsch \& Xing and A.Rasin, which can be clearly understood due to some
different results and statements in their papers \cite{fx} and \cite{rasin},
respectively. Nevertheless, in a pure literal manner, we refer to the
parameterization under consideration as that of Fritzsch-Xing, since their
first paper on the question was submitted for a slightly early time marked in
the referenced journal.}.

To clarify both the presentation of method used for the construction of
model and its physical meaning, we consider the potential for two higgses 
with the discrete $Z_2$-symmetry of vacuum configurations in Section 1. So, the
fermion field has two generations, and the mixing matrix of charged currents
is determined by the only parameter, the Cabibbo angle, which is expressed in
the known way of ``see-saw'' kind \cite{seesaw}. The potential model with three
higgses is constructed in Section 2. The vacuum is symmetric over the
transformations of discrete cyclic group of the third order. This symmetry
provides the introduction of three generations for the fermions. Relations for
the elements of charged-current mixing matrix with the model parameters are
derived. Numerical estimates for the former ones are performed with the use of
current experimental data on the fermion masses and Cabibbo--Kobayashi--Maskawa
matrix. The obtained results are summarized in the Conclusion.

\section{Two generations}

\noindent
Consider a chiral interaction of dirac spinor with an abelian gauge field
${\cal A}_\mu$ and a higgs $h$
\begin{equation}
L=\bar \psi_R p_\mu \gamma^\mu \psi_R +
\bar \psi_L(p_\mu - e{\cal A}_\mu)\gamma^\mu \psi_L +g\bar\psi_R \psi_L h
+g\bar\psi_L \psi_R h^* +L_{GF}+L_0(h,{\cal A}),
\label{f1}
\end{equation}
where $L_{GF}$ is the Lagrangian of free gauge field with a possible
introduction of both a term fixing the gauge and a corresponding Lagrangian for
Faddeev--Popov ghosts, $L_0$ is the higgs Lagrangian with a self-action and the
gauge interaction, $g$ is the Yukawa constant which can be taken real and
positive because of an arbitrary definition of complex phases for the Higgs
field and the product of spinors in the vertex of their interaction.
At the spontaneous breaking of gauge symmetry the vacuum expectation value of
higgs is not equal to zero
\begin{equation}
\langle 0| h |0\rangle = \eta e^{i\phi}.
\label{f2}
\end{equation}
In the unitary gauge $\phi=0$, the fermion gets the mass\footnote{
The appearance of mass for the gauge field is not a subject of consideration
here.} $m=g\eta$
\begin{equation}
L_m = g\eta (\bar \psi_R\psi_L + \bar \psi_L \psi_R).
\label{f3}
\end{equation}
In an arbitrary gauge, it is convenient to use the method of auxiliary fields,
so that, for instance,
\begin{equation}
f_L =\frac{1}{\sqrt{2}} \left( {\psi_L}\atop {e^{i\phi}\psi_L}\right),
\;\;\;\;\;
f_R =\frac{1}{\sqrt{2}} \left( {e^{-i\phi}\psi_R}\atop {\psi_R}\right),
\label{f4}
\end{equation}
and
\begin{equation}
L_m = \bar f_R M_{RL} f_L + \bar f_L M_{LR} f_R,
\label{f5}
\end{equation}
where
\begin{equation}
M_{RL} = M_{LR} = g\eta \left( \begin{array}{cc} 1 & 0 \\ 0 & 1 \end{array}
\right).
\label{f6}
\end{equation}
By the way, the kinetic term of Lagrangian and the gauge interaction of
$f$ with the field $\cal A$ do not change. Thus, one can see from (\ref{f6}),
that the field has the mass $m=g\eta$, as one must expect from the unitary
gauge in (\ref{f3}).

\subsection{Potential of two higgses and generations}

Now consider the analogous interaction of fermions with two higgses
$h_1$ and $h_2$. For the latter ones at the spontaneous breaking of
gauge symmetry, the vacuum expectation values are equal to
\begin{equation}
\langle 0| h_1 |0\rangle = \eta_1 e^{i\alpha}, \;\;\;\;
\langle 0| h_2 |0\rangle = \eta_2 e^{i\alpha+i\phi},
\label{f9}
\end{equation}
where $\alpha$ is an arbitrary gauge parameter,
$\phi$ is the difference between the phases of {\sc vev},
so that this difference can be observable. It is quite evident that the complex
phase leads to the fermion mass depending on $\phi$
\begin{equation}
\frac{1}{g^2} m^2 = (\eta_1+\eta_2\cos\phi)^2+\eta_2^2\sin^2\phi =
(\eta_1+\eta_2)^2-4\eta_1\eta_2\sin^2\frac{\phi}{2}.
\label{f10}
\end{equation}

The following situation is of special interest: The phase difference $\phi$ 
is constrained due to a form of higgs self-action, so that the vacuum
minimum of potential takes place at some discrete values of $\phi$.

In the model under consideration, the higgs potential has the form
\begin{equation}
V=\frac{\lambda_{11}}{4}\big(h_1h_1^*\big)^2-\frac{\mu^2}{2}h_1h_1^*
+\frac{\lambda_{22}}{4}\big(h_2h_2^*\big)^2
-\frac{\lambda_{12}}{8}\big[h_1h_2^*+h_2h_1^*\big]^2,
\label{v}
\end{equation}
where all of $\lambda_{ij}$ and $\mu^2$ are greater than zero, so that the
constraint of the potential stability ($V>0$ at infinity) gives 
$\lambda_{11} \lambda_{22}>\lambda^2_{12}$.

Then, the equations for the energy minimum
$$
\frac{\partial V}{\partial |h_1|} = 0, \;\;\;
\frac{\partial V}{\partial |h_2|} = 0,
$$
result in the {\sc vev} equal to
\begin{eqnarray}
\eta_1^2 & = & \frac{\mu^2}{\lambda_{11}-\frac{\lambda_{12}^2}
{\lambda_{22}}\cos^4 \phi}, 
\label{h1}\\
\eta_2^2 & = & \frac{\lambda_{12}}{\lambda_{22}}\eta_1^2\cos^2 \phi.
\label{h2}
\end{eqnarray}
Then, the minimal energy can be written down in the form
\begin{equation}
V_{min} = -\frac{1}{4}\; \frac{\mu^4}{\lambda_{11}-
\frac{\lambda_{12}^2}{\lambda_{22}}\cos^4\phi}.
\end{equation}
Therefore, in the vacuum state one has $\cos^2 \phi=1$ (see Figs.1 and 2).
Thus, in the specified model of higgs self-action, the phase
difference between the vacuum expectations runs over the discrete values
\begin{equation}
\phi= \pi n,\;\;\; n\in {\rm Z.}
\label{f12}
\end{equation}
It is convenient to use the representation for the cyclic group of second order
$Z_2$, wherein the fields $\{h_1, h_2\}$ have the charges $q=\{0,1\}$,
correspondingly, and the group elements, the relative phases $\exp\{iqn\pi\}$,
are the integer powers of basis $a=\exp\{i\pi\}$. Thus, the configurations
with the various relative phases of {\sc vev} can be obtained by the
transformations of discrete group.

\begin{wrapfigure}[24]{r}{9cm}
\vspace*{3cm}
\hspace*{-0.8cm}
\epsfxsize=6cm \epsfbox{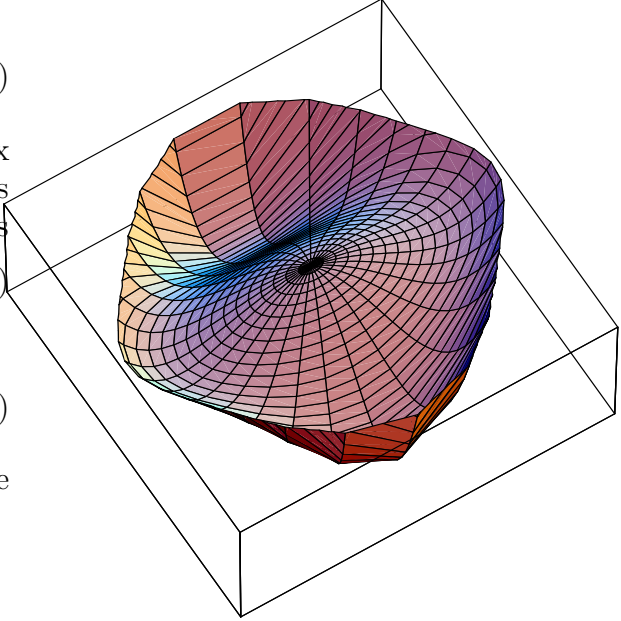}\\

\vspace*{1.3cm}
\caption
{\small \sf The potential of two constant Higgs fields in the model of
(9) at the fixed value of {\sc vev} for the $h_2$ field in accordance
with (11) versus the relative phase $\phi$ and field $|h_1|$ in the polar
coordinates. The measure units are arbitrary,
$\lambda_{11}:\lambda_{12}:\lambda_{22}=10:7:7$.}
\end{wrapfigure}

Then, introducing fields (\ref{f4}), one gets
\begin{equation}
M_{RL} =  g\left( \begin{array}{cc} 
\eta_2 & \eta_1 e^{-2i\phi} \\ \eta_1 & \eta_2 \end{array}
\right),
\label{f11}
\end{equation}
and at fixed $\phi$ (\ref{f12})
the real matrix $M_{RL}=M_{LR}$ has the eigenvalues
corresponding to the fermion masses
\begin{equation}
m_{1,2} = g |\eta_1\pm \eta_2|.
\label{f13}
\end{equation}
Further, introducing
\begin{equation}
U= \frac{1}{\sqrt{2}} \left( \begin{array}{rr} 
1 & 1 \\ -1 & 1 \end{array} \right),
\label{u}
\end{equation} 
one gets the diagonal form of the mass matrix
\begin{eqnarray}
M^U_{LR} &=& M^{U}_{RL} = U \cdot M_{LR}\cdot U^{\dag} 
\nonumber \\
&=& g \left(
\begin{array}{cc} 
\eta_2+\eta_1 & 0 \\ 0 & \eta_2-\eta_1 \end{array} \right)
\nonumber
\end{eqnarray}
with the eigen-vectors $f^{U}_{L,R} = U f_{L,R}$.

Let us emphasize that, first, the introduced auxiliary fields have the 
particular form
$$
f^U_L= \left( \psi^{(1)}_L \atop 0 \right),\;\;\;\;\;
f^U_R= \left( \psi^{(1)}_R \atop 0 \right),
$$
at $\phi=\phi_{(1)}=0$, so that acting on the minimal energy configuration
$|0_{(1)}\rangle$, the field $\psi^{(1)}$ results in the state with the mass
$m^{(1)}= g (\eta_1+\eta_2)$. Second, one obtains
$$
f^U_L= \left( 0 \atop -\psi^{(2)}_L \right),\;\;\;\;\;
f^U_R= \left( 0 \atop \psi^{(2)}_R \right),
$$
at $\phi=\phi_{(2)}= \pi$, so that the $\psi^{(2)}$ field has 
the mass $m^{(2)}= g |\eta_1-\eta_2|$ over the configuration $|0_{(2)}\rangle$.
Therefore, if one constructs the model vacuum as
\begin{equation}
|{\rm vac}\rangle = |0_{(1)}\rangle \otimes |0_{(2)}\rangle =
\left( |0_{(1)}\rangle \atop |0_{(2)}\rangle \right),
\end{equation}
then one can use two kinds of auxiliary fields at $\phi=0$ and $\phi=\pi$,
$f^U$, having the mentioned particular form\footnote{One can easily find that
the auxiliary fields have been chosen in the way, where for each of the given
phase differences the eigen-vectors of mass matrix have been formed.} ~~and the
common mass-matrix independent of the chosen values of $\phi$, {\sf to
introduce} the common {\sf physical} field with the components corresponding to
the kind of generation. So, the {\sf extended}
Lagrangian contains the physical field $f$ over the vacuum $|{\rm vac}\rangle$
with two generations. Note that the vacuum conserves the discrete symmetry of
the second-order cyclic group, which does provide the introduction of fermionic
generations in the given model\footnote{We have defined the number of fermion
fields to be equal to the number of configurations for the vacuum state, i.e.
we have fixed the number of fermionic degrees of freedom in accordance with the
vacuum structure.}.

\begin{wrapfigure}[20]{r}{9cm}
\vspace*{0.4cm}
\hspace*{-0.6cm}
\epsfxsize=7cm \epsfbox{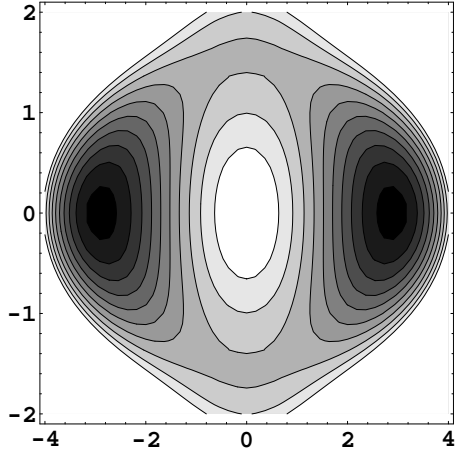}

\vspace*{-0.9cm}
\caption{\small \sf The levels of potential for two constant Higgs fields in
the model of
(9) at the fixed value of {\sc vev} for the $h_2$ field in accordance
with (11) versus the relative phase $\phi$ and field $|h_1|$ in the polar
coordinates. The measure units are arbitrary,
$\lambda_{11}:\lambda_{12}:\lambda_{22}=10:7:7$.}
\end{wrapfigure}

The structure of the Higgs fields vacuum
cannot change the number of fermionic degrees of freedom, so, it cannot be
the origin of generations, since it cannot produce these fermion fields.
Two generations are introduced by construction.
However, the considered model of Higgs vacuum certainly provides
the introduction of two generations: the Higgs vacuum is arranged
for the introduction of two generations of fermions. So, the number of
fermionic degrees of freedom is not changed in an original Lagrangian,
since it is {\sf introduced} in the self-consistent way.

It is correct that in a gauge theory the only possible vacuum
configuration can be chosen because the physical meaning of gauge invariance is
the equivalence of those configurations: The physics in each configuration
is the same, and the only configuration must be chosen, the others are produced
by the gauge transformations. So, the physical principle for the
restriction by the only configuration is the gauge invariance. However, that
is not the case with the model under consideration. The constructed
configurations of minimal energy are not gauge equivalent, since they
distinguish by the phase difference which is the observable physical quantity.
Therefore, the full vacuum of the theory must contain all of the configurations
because there is no physical principle (``a rule of super-selection'') to
cancel some configuration and to restrict the state. So, the paper describes
the theory, where all of the configurations are symmetrically presented in the
vacuum.

The minimal
 energy of potential is symmetric, but the quantum numbers, which are
 the gauge independent differences between the {\sc vev} phases, are changed
 under the action of discrete group. Hence, this symmetry cannot be
 broken spontaneously, since the latter results in forbidding of some 
 values for the quantum numbers, with no physical reasons.
 
The vacuum expectation values are equal to
 $$
 \langle {\rm vac}| h |{\rm vac}\rangle =
 \langle 0_{(1)}| \otimes \langle 0_{(2)}| h  |0_{(1)}\rangle \otimes 
 |0_{(2)}\rangle =
 \left(\begin{array}{cc} \langle 0_{(1)}| h |0_{(1)}\rangle & 0 \\
  0 & \langle 0_{(2)}| h |0_{(2)}\rangle
 \end{array}\right),
 $$
 which is the matrix due to the vacuum definition. So, the higgs {\sc vev},
 which is  a  simple classic field, converts to the matrix with the classic
 elements.  One  can simply recognize that the same representation can be
 equivalently  obtained  by the following construction:
 
 1). Introduce the {\sf replication} of higgs sector for each generation, i.e.
 introduce the matrix for each higgs field
 $$
 h_i \to \left( \begin{array}{cc} h_i^{(1)} & 0\\ 0 & h_i^{(2)}
 \end{array}\right).
 $$
 
 2). Construct the new vacuum as
 $$
 |0\rangle = |0_{(1)}\rangle |0_{(2)}\rangle,
 $$
 where $|0_{(1)}\rangle $ is the single chosen vacuum configuration for the
 first generation with its own higgs sector, and $|0_{(2)}\rangle $ is the
 single vacuum configuration for the second generation with its higgs sector,
 respectively.
 
 Then, the higgs {\sc vev} is the same matrix
 $$
 \langle 0| h |0\rangle =
 \left(\begin{array}{cc} \langle 0_{(1)}| h^{(1)} |0_{(1)}\rangle & 0 \\
  0 & \langle 0_{(2)}| h^{(2)} |0_{(2)}\rangle
 \end{array}\right).
 $$
 Thus, this construction clarifies the situation in the usual way.
 
 The author prefers for the complex definition of the vacuum, when it carries
 an internal  structure, than the replication of higgs sector and, hence, the
 undesirable  breaking of discrete symmetry in each isolated sector, together
 with the  additional observable scalar particles. The first economic way is
 more  attractive to me, but the second one is more evident, simple, usual and
 clear.

In the matrix representation of vacuum state, there is the permutation symmetry
for the vacuum configurations, which is related to the discrete group. These
permutations contain two elements:
$$
S_1= \left(\begin{array}{cc} 1&0 \\ 0&1 \end{array}\right),\;\;\;\;
S_2= \left(\begin{array}{cc} 0&1 \\ 1&0 \end{array}\right),
$$
which correspond to the arbitrary arrangement of generations:
Whether one puts the first generation to be light or heavy, so that in the
initial representation these permutations result in the substitutions
$\eta_1\to \pm\eta_1$ in the mass matrix of (\ref{f11}).

Thus, two species of fermions possessing the identical properties with respect
to the gauge interaction and different mass values, are introduced because of
two kinds of vacuum configurations for the pair of higgses, so that in the
model under consideration, the effect of introduction of two generations is
provided by the vacuum structure of Higgs fields.

From the practical point of view, equation (\ref{f13}) shows
that the large splitting of fermion masses for two generations
($e$, $\mu$) can be caused by no difference between the
values of Yukawa constants for the interaction between the
fermions and higgses. The mass difference can be the result of
small splitting between the vacuum expectation values\footnote{
Supposing $m_1/m_2 = m_e/m_\mu$, one finds that $\Delta\eta/\eta \approx 1/103
\sim \alpha_{em}$, so that, probably, the splitting is of the order of
radiative corrections.} ~~(at $\eta=\eta_1>\eta_2>0$,
$\Delta\eta=\eta_1-\eta_2\ll \eta$)
\begin{equation}
\frac{m_1}{m_2} = \frac{\eta_1-\eta_2}{\eta_1+\eta_2} \approx
\frac{\Delta\eta}{2\eta}\ll 1.
\label{f14}
\end{equation}

Thus, we have shown that at the spontaneous breaking of gauge symmetry
the structure of vacuum for the pair of Higgs fields interacting with the
fermions can provide the introduction of two generations of fermions as two
kinds of the relation with the vacuum symmetric over the discrete group.

\subsection{Mixing matrix. Cabibbo angle.}

The introduction of auxiliary fields (\ref{f4}) with the symmetric mass-matrix
at $\phi=0, \pi$ contains the uncertainty related to an additional term 
$$
\Delta M = \left( \begin{array}{rc} -a & 0\\ 0 & a \end{array} \right),
$$
which does not contribute to the determination of the mass values,
since it is canceled in the Lagrangian expressed through the initial
single-generation fields. This uncertainty corresponds to the rotation of the
{\sf introduced} physical fields in the Lagrangian with two generations.
So, considering the general form of extended Lagrangian with different Yukawa
constants $g_1$ and $g_2$ for the Higgs fields $h_1$ and $h_2$, one gets the 
following expression for the mass matrix\footnote{The number of relative
phases, which cannot be removed due to redefinitions in the set of couplings
$\{g_1,g_2,\ldots,g_n\}$ and fields $\{h_1,h_2,
\ldots,h_n\}$, is equal to $n_\delta=(n-1)(n-2)/2$, so that for $n=2$,
we have the situation, when $n_\delta=0$, and $\{g_1,g_2\}$ are real and
positive.}~~:
\begin{equation}
M = \left( \begin{array}{cc} v_2-v_1 \sin 2\theta & v_1 \cos 2\theta \\
v_1 \cos 2 \theta & v_2+v_1 \sin 2 \theta \end{array} \right),
\end{equation}
where $v_i=g_i\eta_i$, so that the eigenvalues of the matrix are the same 
$m_{1,2}=|v_1\pm v_2|$, which are independent of the $\theta$ value. However,
there is a particular state, when $\theta$ is determined by the
physical quantities of model. So, we find\footnote{We consider the case of
$v_2/v_1\le 1$. A description of the inverse condition is quite evident after 
a suitable transformation of $\psi_R$.}
\begin{equation}
M = \left( \begin{array}{cc} v-a & v\\ v & v+a \end{array} \right),
\label{mv}
\end{equation}
where $v=v_2$, $a=v \tan 2 \theta$, 
$$
\cos 2\theta = \frac{v_2}{v_1}= \frac{g_2}{g_1}
\sqrt{\frac{\lambda_{12}}{\lambda_{22}}}.
$$ 
The $\theta$ value is related to the masses of fermions
\begin{equation}
\tan 2\theta = 2 \frac{\sqrt{m_1 m_2}}{m_2-m_1}.
\end{equation}
The matrix gets the ``see-saw'' form in the ``heavy'' basis \cite{dem}
\begin{equation}
M^{U} = U\cdot M\cdot U^{\dag} = \left( \begin{array}{cc} 2v & a\\ a & 0
\end{array}
\right),
\label{mv2}
\end{equation}
where $U$ is defined in eq.(\ref{u}). At $v_1=v_2$ we have $a=0$, and 
the mass matrix of (\ref{mv}) is a ``democratic'' one \cite{dem}.

Next, the model matrix of (\ref{mv2}) takes the diagonal form after
the action by the rotation to the angle $\theta$.

Let us consider now the model with two kinds of the chiral fields 
$\psi^{(u)}$ and $\psi^{(d)}$ possessing different charges, so that
the charged current interaction between the latter ones has the form
$$
L_{cc} = e \bar \psi^{(u)}_L W_\mu\gamma^\mu \psi^{(d)}_L + \mbox{h.c.},
$$
where the initial mass-matrices have the form of (\ref{mv}) with
$$
\cos 2\theta^{(u)}
=\frac{g_2^{(u)}}{g_1^{(u)}}\sqrt{\frac{\lambda_{12}}{\lambda_{22}}},\;\;\;
\cos 2\theta^{(d)}
=\frac{g_2^{(d)}}{g_1^{(d)}}\sqrt{\frac{\lambda_{12}}{\lambda_{22}}}.
$$
Then, the Cabibbo mixing matrix has the form
$$
V = \left( \begin{array}{rc} \cos \theta_c & \sin \theta_c \\
-\sin \theta_c & \cos \theta_c \end{array} \right),
$$ 
where $\theta_c=\theta^{(u)}-\theta^{(d)}$. At $\frac{m_2^{(u)}}{m_1^{(u)}}\ll
\frac{m_2^{(d)}}{m_1^{(d)}}\ll 1$, one gets the Cabibbo angle
$$
\sin\theta_c\approx \; \sqrt{\frac{m_1^{(d)}}{m_2^{(d)}}}.
$$
Thus, the offered model provides the introduction
of two generations due to the structure of vacuum for
two Higgs fields. It describes also the mixing matrix of charged 
currents through the physical quantities, which allows one to relate 
the mixing angle with the masses of fermions.

\section{Three generations}

Constructing the model in this section, we will follow the scheme presented
above in the case of two generations.

Consider the potential of three higgses
\begin{eqnarray}
V &=& \frac{\lambda_{11}}{4} (h_1^* h_1)^2 - \frac{\mu^2}{2} h_1^* h_1 +
\frac{\lambda_{22}}{4} [(h_2^* h_2)^2+(h_3^* h_3)^2]-
\nonumber\\ &&
\frac{\lambda_{12}}{8}[h_1^* h_2+h_2^* h_1+h_1^* h_3+h_3^* h_1+
h_2^* h_3+h_3^* h_2-
\nonumber\\ &&
\frac{k}{2}(h_1^* h_1+h_2^* h_2+h_3^* h_3)]^2.
\label{v3g}
\end{eqnarray}
As it has been mentioned in the previous section, the relative phases in the
sum of pair products of three higgses in the expression for their potential can
be removed by the redefinitions of fields so that there are no additional
complex phases in the second string of (\ref{v3g}). Moreover, in the contact
terms of higgs interaction with the fermion field one can make two quantities
of three Yukawa constants to be positive by the same procedure. So, there is
the only complex parameter, a single Yukawa constant.

The vacuum configuration corresponds to constant fields determined by five
parameters: three absolute values of {\sc vev} $\eta_i$ and two phase
differences between them, $\phi_j={\rm arg}[h_j h_1^{*}]$. Putting the first
derivatives of potential over these quantities to be equal to zero, we find
$$
V_{min}= -\frac{\mu^2}{4} \eta_1^2
$$
from the condition $\sum \eta_i\cdot \frac{\partial V}{\partial \eta_i}=0$.
Further, the derivatives over the phase differences give
\begin{eqnarray}
\eta_1 \sin(\phi_1)+\eta_3 \sin(\phi_1-\phi_2) &=& 0,\\
\eta_1 \sin(\phi_2)-\eta_2 \sin(\phi_1-\phi_2) &=& 0,
\end{eqnarray}
for nonzero values of $\eta_i$. A consideration of $\phi_j=\{0,\pi\}$
can be easily performed, so that we study the configuration, wherein the latter
ones are excluded. Then, we get
\begin{eqnarray}
\phi_1 &=& - \phi_2,\\
\cos^2\phi_1 &=& \frac{2\lambda_{22}}{\lambda_{12}(2+k)^2}- \frac{1}{2},\\
\eta_1^2 &=& \frac{\mu^2}{\lambda_{11}-\frac{\lambda_{12}}{16 \cos^2 \phi_1}
(4+4k+k^2)(2\cos^2\phi_1+1)}.
\end{eqnarray}
In the rest of paper we use the following set of potential parameters
\begin{eqnarray}
k &=& 1, \\
\lambda_{11} &=& 20 \lambda_{12},\\
\lambda_{22} &=& \frac{27}{8} \lambda_{12},
\end{eqnarray}
which leads to
\begin{eqnarray}
&&\eta_1 =\eta_2 = \eta_3, \\
&&\phi_1= \pm \frac{2\pi}{3},
\end{eqnarray}
where we have taken into account the condition $\eta_i >0$. For the chosen set
of parameters the same absolute values of both {\sc vev} and minimal energy are
reached at $\phi_j=0$, too.

Thus, in the described model of potential for three Higgs fields the vacuum
configuration corresponds to three possible sets of phase differences:
$\phi_1=\{0;\pm 2\pi/3\}$, $\phi_2=-\phi_1$. The fact means that the model
vacuum is symmetric over the action of third-order cyclic group
$Z_3$, whereas the fields $\{h_1,h_2,h_3\}$ possess the charges
$\{0,1,-1\}$, correspondingly. As it was in the case of two generations,
it is convenient to use the matrix representation for the vacuum state defined
as the direct product of three given configurations of minimal energy
\begin{equation}
|{\rm vac}\rangle = |0_{(1)}\rangle \otimes |0_{(2)}\rangle \otimes
|0_{(3)}\rangle=
\left( \begin{array}{c} |0_{(1)}\rangle \\  |0_{(2)}\rangle \\
|0_{(3)}\rangle \end{array}\right),
\end{equation}
or in the similar way, we can introduce the replication of higgs sector for
each of three generations and define the vacuum state as
\begin{equation}
|0\rangle = |0_{(1)}\rangle |0_{(2)}\rangle |0_{(3)}\rangle,
\end{equation}
where we have already chosen the single vacuum configuration for the each
sector of higgses, so that the same matrix of {\sc vev} is reproduced, when we
take the model with $|{\rm vac}\rangle$.

Acting on each element of vacuum $|{\rm vac}\rangle$, the fermionic operator
produces the field with different masses: three configurations lead to three
values of masses. Thus, the discrete symmetry of vacuum provides the
introduction of three generations of fermionic field.

In the Lagrangian for the interaction of higgses with the fermion field
\begin{equation}
L_m=g_1\bar\psi_R \psi_L h_1+g_1\bar\psi_L \psi_R h_1^*+
g_2\bar\psi_R \psi_L h_2+g_2\bar\psi_L \psi_R h_2^*+
g_3\bar\psi_R \psi_L h_1+g_3\bar\psi_L \psi_R h_3^*,
\label{lm}
\end{equation}
we will suppose that $g_{2,3}>0$, and
$$
g_1 = |g_1| e^{i\epsilon}.
$$
Introduce the notations
\begin{eqnarray}
g_1 \cdot \eta_1 & = & v+i\cdot \bar v,\\
g_2 \cdot \eta_2 & = & v_2,\\
g_3 \cdot \eta_3 & = & v_3,
\end{eqnarray}
and 
\begin{eqnarray}
v' & = & \frac{1}{2}(v_2+v_3),\\
\bar v' & = & \frac{\sqrt{3}}{2}(v_2-v_3).
\end{eqnarray}
Then the masses of generations are determined by the expressions
\begin{eqnarray}
m_3^2 & = & (v+2v')^2+\bar v^2,\\
m_2^2 & = & (v-v')^2+(\bar v+\bar v')^2,\\
m_1^2 & = & (v-v')^2+(\bar v-\bar v')^2.
\end{eqnarray}
Construct the auxiliary fields
\begin{equation}
f_L = \frac{1}{\sqrt{3}}
\left( \begin{array}{c} \psi_L \\ e^{-2i\phi_1} \psi_L \\ 
e^{-2i\phi_2} \psi_L \end{array}\right),\;\;\;
f_R = \frac{1}{\sqrt{3}}
\left( \begin{array}{c} \psi_R \\ e^{-i\phi_1} \psi_R \\ 
e^{-i\phi_2} \psi_R \end{array}\right).
\label{a3}
\end{equation}
Then the Lagrangian of (\ref{lm}) can be written down in terms of fields
(\ref{a3}) with the use of mass matrix
\begin{equation}
\tilde M_{RL} = \left( \begin{array}{ccc} 
v_1 & v_2 e^{3i\phi_1} & v_3 e^{3i\phi_2}\\
v_2 & v_3 e^{i(\phi_1+\phi_2)} & v_1 e^{i(2\phi_2-\phi_1)}\\
v_3 & v_1 e^{i(2\phi_1-\phi_2)} & v_2 e^{i(\phi_1+\phi_2)} 
\end{array}\right).
\end{equation}
In the considered model of potential the admissible set of phase differences
results in the mass matrix, which does not depend on the choice of vacuum
configuration
\begin{equation}
M_{RL} = \left( \begin{array}{ccc} 
v_1 & v_2 & v_3 \\
v_2 & v_3 & v_1 \\
v_3 & v_1 & v_2  
\end{array}\right).
\label{m3}
\end{equation}
In a general case, matrix (\ref{m3}) is not hermitian because of the complex
value of $v_1$. It possesses the permutation symmetry over the indices of
higgses in the way that its eigenvalues do not change through these
permutations. The elements of the permutation group are given by the matrices
\begin{eqnarray}
S_1=\left({\begin{array}{ccc} 
1 & 0 & 0\\ 0 & 1 & 0\\ 0 & 0 & 1\end{array}}\right), \; & \; &
S_2=\left({\begin{array}{ccc} 
0 & 1  & 0\\ 1 & 0 & 0\\ 0 & 0 & 1\end{array}}\right), \\
S_3=\left({\begin{array}{ccc} 
1 & 0 & 0\\ 0 & 0 & 1\\ 0 & 1 & 0\end{array}}\right), \; & \; &
S_4=\left({\begin{array}{ccc} 
0 & 0 & 1\\ 0 & 1 & 0\\ 1 & 0 & 0\end{array}}\right).
\end{eqnarray}
The mass matrix can be diagonalized by the action of  unitary matrices
$$
U_R={1 \over \sqrt{3}}\pmatrix{ 
1 & 1 & 1 \cr
1 & \omega & \omega^2 \cr
1 & \omega^2 & \omega \cr},\;\;\;
U_L^{\dagger}={1 \over \sqrt{3}}\pmatrix{
1 & 1 & 1 \cr
1 & \omega & \omega^2 \cr
1 & \omega^2 & \omega \cr},
$$
where $\omega=\exp\{i2\pi/3\}$, so that
$$
U_R M_{RL} U_L^{\dagger}=\pmatrix{ 
M_3 & 0 & 0 \cr
0  & M_2 & 0 \cr
0  & 0 & M_1 \cr},
$$
where $|M_i|=m_i$,
\begin{eqnarray}
M_3 &=& v_1+v_2+v_3,\\
M_2 &=& v_1+v_2\omega+v_3\omega^2,\\
M_1 &=& v_1\omega^2+v_2\omega+v_3.
\end{eqnarray}
Further, note that the unitary matrices $U^{\dagger}$ diagonalizing
$M_{RL}$, are constructed from the corresponding vectors, which are given by
the auxiliary fields of (\ref{a3}) at three fixed phase differences of higgs
{\sc vev}. The fact means that, as it was in the model with two generations,
each configuration of auxiliary field in the diagonal representation
corresponds to the only generation, exactly, i.e. the only element of vector is
not equal to zero. Thus, we can {\sf introduce} the physical field $f$, whose
components correspond to the generations of fermion field, with the common
mass-matrix of (\ref{m3}).

\subsection{Mixing matrix of charged currents. Phase of {\sf CP}-violation}

Having considered the example of two fermionic generations, we see that
the initial mass-matrix for the auxiliary fields contains the uncertainty,
which does not change the values of masses and can be related to the
rotations of components for the extended physical field. For three generations,
a displacement of relative phase between the generations is added to the
rotations. The rotation angles and the complex phase cannot be any external
additional parameters of the model, and they have to be related to the
physical quantities: $v,v',\bar v, \bar v'$. So, the transformation parameters
can be determined due to a fixing of a symmetric form of the mass matrix. For
the system of two generations, the symmetry condition for the $2\times 2$
matrix has looked like:
\begin{equation}
M(1,2)=M(2,1), \;\;\;\; \frac{1}{2}[M(1,1)+M(2,2)]=M(1,2),
\end{equation}
which does not depend on the mentioned permutation symmetry for the
components. The symmetry condition leads to definite relations of
the Cabibbo angle with the physical parameters of model. The natural
unification of condition to the case of three generations is given by the
following equations:
\begin{eqnarray}
& M(1,2)=M(2,1),\;\; M(1,3)=M(3,1),\;\; M(2,3)=M(3,2),\\
& \frac{1}{2}[M(1,2)+M(2,3)]=M(1,3),
\label{cond}
\end{eqnarray}
so that the resulting matrix has the form
\begin{equation}
M = \left(\begin{array}{ccc} 
\mu_1 & \bar\mu+\Delta & \bar\mu \\
\bar\mu+\Delta & \mu_2 & \bar\mu-\Delta \\
\bar\mu & \bar\mu-\Delta & \mu_3 \end{array}\right).
\label{m-cond}
\end{equation}
It can be constructed in the following way:

Permute elements $v_1$ and $v_2$ in matrix (\ref{m3}) (the action of
permutation operator $S_4$) and come to the ``heavy'' basis by the
transformation
$$
U\cdot M_{RL}\cdot U^{\dagger} = M_U,
$$
where
\begin{equation}
U = \left(\begin{array}{ccc} 
1/\sqrt{3} & 1/\sqrt{3} & 1/\sqrt{3} \\
1/\sqrt{2} & -1/\sqrt{2} & 0 \\
1/\sqrt{6} & 1/\sqrt{6} & -2/\sqrt{6} \end{array}\right),
\label{us}
\end{equation}
so that
\begin{equation}
M_U = \left(\begin{array}{ccc}
v+2v'+i\cdot \bar v & 0 & 0 \\
0 & v'-v-i\cdot\bar v & \bar v' \\
0 & \bar v' & v-v'+i\cdot\bar v \end{array}\right).
\end{equation}
Further, transform this matrix to the form
\begin{equation}
\tilde M_U = \left(\begin{array}{ccc}
v+2v'+i\cdot \bar v & 0 & 0 \\
0 & -\bar v+i\cdot(v-v') & \bar v' \\
0 & \bar v' & -\bar v+i\cdot(v-v') \end{array}\right)
\label{m-t}
\end{equation}
by the action of operators
$$
U_1 = \left(\begin{array}{ccc}
1 & 0 & 0 \\
0 & -i & 0 \\
0 & 0 & 1 \end{array}\right),\;\;
U_2 = \left(\begin{array}{ccc}
1 & 0 & 0 \\
0 & 1 & 0 \\
0 & 0 & i \end{array}\right)
$$
so that $\tilde M_U = U_2\cdot M_U\cdot U_1$. As one can see, matrix
(\ref{m-t}) has the form similar to one considered above in the case of two
generations, so that the third generation is decoupled, and the real part
of matrix for two other generations exactly corresponds to the form of matrix
constructed above. By the way, an imaginary part of matrix for two generations
is proportional to the identical one. Then, we can easily introduce the
rotation
in an analogous way as in the model with two generations, so that
\begin{equation}
\cos 2\theta_{12} = \frac{\bar v}{\bar v'},
\end{equation}
and the mass matrix is transformed to 
$$
M_{(12)} = S_{(1)}^{\dagger}\cdot \tilde M_U \cdot S_{(1)},
$$
where
$$
S_{(1)} = \left(\begin{array}{ccc}
1 & 0 & 0 \\
0 & \cos \theta_{12} & \sin \theta_{12} \\
0 & -\sin \theta_{12} & \cos \theta_{12} \end{array}\right).
$$
Further, make the inverse transformation with matrices $U_{1,2}$ to
get the previous texture of imaginary parts for the elements of mass matrix:
$\tilde M_{(12)} = U_2^{\dagger}\cdot M_{(12)}\cdot
U_1^{\dagger}$. Then, introduce both the angle of mixing with the third (heavy)
generation and the relative phase
$$
S_{(3)} = \left(\begin{array}{ccc}
\cos \theta_{3} & 0 & \sin \theta_{3} \\
0 & e^{i\delta} & 0 \\
-\sin \theta_{3} & 0 & \cos \theta_{3} \end{array}\right),
$$
whereas  
$$
M_{(3)} = S_{(3)}^{\dagger}\cdot \tilde M_{(12)} \cdot S_{(3)}.
$$
In what follows we will use the approximation of low $\theta_3$ value in the
linear order over this angle. Finally, coming from the ``heavy'' basis to the
``democratic'' one, we find the mass matrix $M$
$$
M = U^{\dagger}\cdot M_{(3)}\cdot U.
$$
Making conditions (\ref{cond}) to be valid, we get that the mass matrix takes
form (\ref{m-cond}), if the following equations are satisfied
\begin{eqnarray}
-\frac{2}{3\sqrt{2}}\; \frac{v-v'+\sqrt{3} \bar v \cos \delta}
{v' +\frac{1}{\sqrt{3}} \bar v \cos \delta} & = & \theta_{3}, \\
6 - 8\sqrt{3} \sin \delta - \sqrt{6} \theta_3 \sin \delta +
3\sqrt{2} \theta_3 \tan 2\theta_{12} & = & 0.
\end{eqnarray}
Constructing a realistic model, we have to use the condition of observed
low values of masses for the junior generations with respect to the mass of
major generation as well as the fact that the mixing between two junior
generations is quite accurately described by the scheme considered in the model
of two generations. So, in the realistic model we suppose
$$
\begin{array}{ccc}
|v-v'| & \ll & |\bar v|, \\
|\bar v| & \ll & v,v', \\
|v-v'| & \ll & |\bar v-\bar v'|.
\end{array}
$$
Then we can easily find
\begin{equation}
\begin{array}{ccl}
\sin\delta & \approx & \frac{\sqrt{3}}{4},\\
\theta_3 & \approx & -\sqrt{\frac{3}{2}} \frac{2\bar v}{3 v'}\cos\delta 
\approx \pm\sqrt{\frac{39}{32}} \frac{m_2}{m_3},\\
\tan 2\theta_{12} & \approx & 2 {\sqrt{m_1 m_2}}/{(m_2-m_1)},\\
m_3 & \approx & 3 v,\\
m_2 & \approx & |\bar v|+|\bar v'|,\\
m_1 & \approx & ||\bar v|-|\bar v'||.
\end{array}
\end{equation}
First of all, note that at small angles of generation mixing, the introduction
of complex phase $\delta$ is necessary, whereas the value of $\sin\delta$ 
in the leading approximation is certainly fixed. However, we have to point out
such uncertainties as signs of $\cos\delta (\delta\to \pi-\delta)$, $\bar v$,
$\bar v'$, and, hence, signs of some masses and rotation angles. Therefore,
the order of arrangement for the first and second generations is not definite
in the diagonal representation, since the relative sign of $\bar v$ and $\bar
v'$ is uncertain. The phenomenological analysis shows that the angle $\theta_3$
corresponds to the mixing between the second generation and the third one.

Now consider the picture of mixing for the charged currents of left-handed
fermions, $\psi^{(u)}$ and $\psi^{(d)}$. One can easily understand that
the unitary matrix of Cabibbo--Kobayashi--Maskawa is given by the
expression\footnote{In the model with two generations the additional
transformation by matrix (\ref{u}) does not influence the Cabibbo
matrix because this transformation is the rotation to the angle
$\frac{\pi}{4}$, which is compensated by the inverse rotation for the quarks of
other charge. In the model with three generations, the presence of mixing with
the third heavy generation leads to the fact that to remove the additional
rotation to the angle $\frac{\pi}{4}$, we have correspondingly to transform the
mass matrix: to come to the ``heavy'' basis, to make the inverse rotations by
matrix $S_{(3)}$ and operators $U_{1,2}$, to rotate the matrix of two light
generations by  (\ref{u}), and, finally, to be back to the initial texture
(operators $U_{1,2}$, $S_{(3)}$ and the rotation from the ``heavy'' basis to
the ``democratic'' one (\ref{us})). By the procedure, the mentioned properties
of symmetry for the mass matrix naturally survive in a changed form.}
$$
V_{CKM} = S_{(1)}^{(u)}\cdot S_{(3)}^{(u)}\cdot S_{(3)}^{(d)\dagger}\cdot
S_{(1)}^{(d)\dagger}.
$$
The complex phases in the matrices $S_{(3)}$ for the ``up'' kind and ``down''
kind fermions do not cancel each other, if $\delta^{(u)}=\pi-\delta^{(d)}$. For
the mixing angles of junior generations $\theta^{(u,d)}=\theta_{12}^{(u,d)}$
with the accuracy up to the sign, we approximately get 
$$
\sin \theta^{(u,d)} \approx \pm\sqrt{\frac{m_1^{(u,d)}}{m_2^{(u,d)}}}.
$$
Thus, the mixing matrix $V_{CKM}$ in the representation of Fritzsch--Xing has
the form
$$
V_{CKM} = \left(\begin{array}{ccc}
s_u s_d c + c_u c_d e^{-i\tilde\delta} & s_u c_d c - c_u s_d e^{-i\tilde\delta}
& s_u s \\
c_u s_d c - s_u c_d e^{-i\tilde\delta} & c_u c_d c + s_u s_d e^{-i\tilde\delta}
& c_u s \\
-s_d s & -c_d s & c 
\end{array} \right),
$$
where $\cos \theta^{(u,d)} = c_{u,d}$, and analogous definitions are made for
the sines, and
$$
\begin{array}{ccl}
\tilde \delta & \approx & \pi \pm 2 \arcsin \frac{\sqrt{3}}{4},\\
\theta & = & \theta_3^{(u)}-\theta_3^{(d)}.
\end{array}
$$
The features of such representation of the mixing matrix for the charged
currents are in detail discussed in papers by Fritzsch and Xing.
Let us here note only that with the accuracy up to small corrections over
the mass ratios for the fermion generations, we have
\begin{eqnarray}
\cos \tilde \delta &=& -\frac{5}{8}, \label{cp2}\\
|\theta| &=& \sqrt{\frac{39}{32}}\; \bigg|\frac{m_2^{(u)}}{m_3^{(u)}}\pm
\frac{m_2^{(d)}}{m_3^{(d)}}\bigg|.
\end{eqnarray}

\subsection{Numerical estimates}
Consider the matrix of charged current mixing for quarks. This matrix is of
special interest phenomenologically, since the measurement of its elements is
carried out experimentally in the framework of precision study of the Standard
Model, but also theoretically because, in contrast to lots of models
introducing the various initial mass-matrices of the fermion generations, the
model under consideration can make some definite predictions on both the phase
of {\sf CP}-violation (\ref{cp2}) and the relation of angle $\theta$ with the
masses of quarks.

Indeed, with the accuracy up to small corrections linear over the ratios of
generation masses, we find
$$
|V_{cb}|\approx|\theta| = \sqrt{\frac{39}{32}}\; 
\bigg|\frac{m_s}{m_b}\pm \frac{m_c}{m_t}\bigg|.
$$
For the Cabibbo angle, we get
$$
\sin \theta_c = |s_u-s_d  e^{-i\tilde\delta}| \approx
|s_d+s_u \cos \tilde\delta| = \bigg|\sqrt{\frac{m_d}{m_s}}\pm
\frac{5}{8}\;\sqrt{\frac{m_u}{m_c}}\bigg|,
$$
where the upper sign corresponds to the negative value of relative sign between 
$s_u$ and $s_d$. Further\footnote{See a discussion of such relations in
\cite{rasin2,fx2}.},
\begin{eqnarray}
\bigg| \frac{V_{ub}}{V_{cb}}\bigg|  &=& \tan \theta_u = 
\sqrt{\frac{m_u}{m_c}},\\
\bigg| \frac{V_{td}}{V_{ts}}\bigg|  &=& \tan \theta_d = 
\sqrt{\frac{m_d}{m_s}}\approx \sin\theta_c.
\end{eqnarray}
In the $B$-meson physics, angles of unitary triangle are usually introduced.
So, in the given model we can predict, for example,
$\alpha={\rm arg}[-V_{td}V_{tb}^*V_{ud}^*V_{ub}]$, which approximately
coincides with the angle $\tilde\delta$ up to high accuracy,
$$
\sin 2\alpha\approx \sin 2\tilde\delta =\pm \frac{5}{32}\sqrt{39}\approx
\pm 0.976.
$$
For $\beta={\rm arg}[-V_{cd}V_{cb}^*V_{td}^*V_{tb}]$, we obtain
$$
\sin 2\beta \approx 2\frac{s_u}{s_d} \sin\tilde\delta\; 
\bigg[1+\frac{s_u}{s_d} \cos\tilde\delta\bigg]
\approx \pm\frac{\sqrt{39}}{4}\; \sqrt{\frac{m_u m_s}{m_d m_c}}
\bigg[1\mp\frac{5}{8}\sqrt{\frac{m_u m_s}{m_d m_c}}\bigg].
$$
The universal parameter ${\cal J}$ \cite{j} independent of the representation
for the mixing matrix of charged currents determines the violation of
{\sf CP}-invariance. In the model it is given by the expression
$$
{\cal J} = s_uc_us_dc_d s^2 c \sin\tilde\delta
\approx \pm \frac{39}{32}\;\bigg|\frac{m_s}{m_b}\pm\frac{m_c}{m_t}\bigg|^2
\sqrt{\frac{m_u m_d}{m_s m_c}}\frac{\sqrt{39}}{8}.
$$
Choosing the negative sign in the expressions for the Cabibbo angle and element
$V_{cb}$ and supposing that\footnote{In this paper we are not concerned with
the problem on the values of quark masses: a scale for the determination of
``running'' masses, a relation of the current masses with the pole ones,
uncertainties of theoretical models in QCD for an extraction of the quark
masses from the experimental data and so on. Note only that an admissible
variation of quark masses for two light generations can reach 50\%, 
because of both effects of QCD evolution to a scale of the order of
$t$-quark mass and some uncertainties in the initial data of the evolution.
So, the set supposed in (\ref{maset}) is generally used for the illustrative
purpose though it is quite typical.}
\begin{equation}
\begin{array}{crlcrl}
m_d= & 14 & {\rm MeV}, & m_u= & 7 & {\rm MeV,}\\
m_s= & 200 & {\rm MeV}, & m_c= & 1.4 & {\rm GeV,}\\
m_b= & 4.6 & {\rm GeV}, & m_t= & 175 & {\rm GeV,}
\end{array}
\label{maset}
\end{equation}
we find\footnote{Note that the admissible variation of quark masses results in
valuable uncertainties in the determination of angle $\beta$ 
($\Delta \beta \sim 5\%$) and elements of mixing matrix. We have also to point
out that the choice of quark masses close to the values in the paper by Lucas
\& Raby \cite{super} leads to the quark-currents mixing matrix approximately
equal to what is obtained in the model under consideration.} ~~that
$$
|\sin 2\beta |= 0.490,\;\;\; |{\cal J}|\approx 2.2\cdot 10^{-5},
$$
and the set of values shown in Table \ref{vckm} for the elements of
Cabibbo--Kobayashi--Maskawa matrix.

\begin{table}[th]
\caption{The model predictions with the mass set of (67) 
in comparison with the current experimental data.}
\begin{center}
\begin{tabular}{|c|l|l|}
\hline
quantity & model & experiment \\
\hline
$|V_{us}|$ & $0.220$ & $0.2205\pm 0.0026$ \\
$|V_{cb}|$ & $0.0392$ & $0.0392\pm 0.003$ \\
$|V_{ub}/V_{cb}|$ & $0.071$ & $0.08\pm 0.02$ \\
$|V_{td}/V_{ts}|$ & $0.265$ & $0.22\pm 0.07$ \\
\hline
\end{tabular}
\end{center}
\label{vckm}
\end{table}

Thus, we have shown that in the framework of model the numerical estimates of
matrix elements for the mixing of charged quark currents agree with the current
experimental data within the accuracy of determination, but also they have a
predictive power with respect to the parameters of {\sf CP}-violation.

\section*{Conclusion}

In this paper we have constructed the models of potentials for two and three
higgses, wherein there are the discrete symmetries of vacuum configurations,
$Z_2$ and $Z_3$, correspondingly, which provides the introduction of two and
three generations of fermions coupled to the higgses by the Yukawa interaction.
Thus, the essential reason for the generation replication is in part the
symmetric structure of higgs vacuum.

In the framework of such approach, the hierarchy of generation masses
(two generations are light and the only one is heavy) is caused by a small
splitting of the quantities defined by the products of vacuum expectation
values of higgses and Yukawa coupling constants in the higgs-fermion contact
terms.

These physical parameters of model determine the matrix of charged current
mixing, whose elements are related to the ratios of generation masses.
In the leading approximation for the quark currents mixing, the hierarchy of
masses results in the certain predictions on the complex phase of
{\sf CP}-violation and several elements of the matrix, though there is the
uncertainty in the choice of several signs because of indefinite values for the
model parameters: the arrangement of Yukawa constants and vacuum expectation
values. We have shown that the typical set of quark masses leads to the
Cabibbo--Kobayashi--Maskawa matrix which is in a good agreement with the
current experimental data in the framework of error bars.

{\hfill \it Received February 24, 1998}
\end{document}